\documentclass[conference]{IEEEtran}
\IEEEoverridecommandlockouts
\usepackage[english]{babel}
\usepackage[utf8x]{inputenc}
\usepackage[T1]{fontenc}

\usepackage{amsmath}
\usepackage{graphicx}
\usepackage[colorinlistoftodos]{todonotes}
\usepackage[colorlinks=true, allcolors=blue]{hyperref}
\usepackage{multirow}
\usepackage{rotating}

\usepackage{enumitem}
\setlist[enumerate]{label=\roman*}
\usepackage{wrapfig,lipsum,booktabs}
\usepackage{placeins}
\usepackage{float}
\usepackage{enumitem}
\usetikzlibrary{mindmap}
 \usepackage[flushleft]{threeparttable}
\usepackage{setspace}
\usepackage{multicol}
\usepackage{supertabular}
\usepackage{wasysym}   
\usepackage{longtable}
\usepackage{tikz}
\usetikzlibrary{shapes.geometric, arrows,positioning}
\usetikzlibrary{shapes,arrows,positioning,patterns,decorations.pathmorphing}
\usepackage[linguistics]{forest}
\usepackage{adjustbox}
\usepackage{array}
\usepackage{booktabs}
\newcommand*\rot{\rotatebox{90}}
\usepackage{pifont}

\title{Evaluating Password Advice}
\author{\IEEEauthorblockN{Hazel Murray}
\IEEEauthorblockA{Department of Mathematics and Statistics\\Maynooth University, Ireland
\\Email: hazelmsmurray@gmail.com\thanks{This publication has emanated from research supported in part by a research grant from Science Foundation Ireland (SFI) and is co-funded under the European Regional Development Fund under Grant Number 13/RC/2077. This research is also supported by a John and Pat Hume doctoral studentship.}}
\and
\IEEEauthorblockN{David Malone}
\IEEEauthorblockA{Hamilton Institute\\Maynooth University, Ireland\\
E-mail: david.malone@nuim.ie
}}

\begin{document}
\maketitle

\begin{abstract}
Password advice is constantly circulated by standards agencies, companies, websites and specialists. But there appears to be great diversity in terms of the advice that is given. Users have noticed that different websites are enforcing different restrictions. For example, requiring different combinations of uppercase and lowercase letters, numbers and special characters. We collected password advice and found that the advice distributed by one organization can directly contradict advice given by another. Our paper aims to illuminate interesting characteristics for a sample of the password advice distributed. We also create a framework for identifying the costs associated with implementing password advice. In doing so we identify a reason for why password advice is often both derided and ignored. 
\end{abstract}

\section{Introduction} 
Password advice is important for informing users and organizations about how to best maintain high standards of security. However, the content of the advice given is often not compatible across different sources. Our paper aims to understand the composition of the advice circulated to users and organizations and to identify key costs associated with the implementation of this advice. We collected a large selection of password advice and categorized it in order to highlight characteristics and discrepancies. Using the collected advice we were able to identify costs linked with implementing the advice. We believe this provides insight into why users and organizations are not enforcing researcher’s recommended password practices. 

In his 2009 paper, Herley \cite{herley2009so} argues that users' rejection of security advice is rational from an economic perspective. Herley identifies advice relating to password security, phishing and certificate errors. For each, he discussed the costs and the potential/actual benefits. We will build on Herley's work through our creation of a framework for identifying the costs associated with enforcing password advice. 

Despite the wide distribution of advice and the general acknowledgement of inconsistencies \cite{bonneau2010password}, a framework for simple analysis of advice is not available. Yet, many researchers have identified problems with the advice given \cite{zhang2016revisiting}. Inglesant and Sasse \cite{inglesant2010true} find that users are, in general, concerned with maintaining security but that existing security policies are too inflexible to match their capabilities, and the tasks and contexts in which they operate. We hope to develop this idea by identifying some of the demands associated with following passwords advice. Flor{\^e}ncio, Herley and van Oorschot \cite{florencio2014password} find that mandating exclusively strong passwords with no reuse gives users an effectively impossible task as portfolio size grows. Bellovin \cite{bellovin2008security} questions whether simple adherence to password advice on security checklists really accomplishes the desired security goals. Flor{\^e}ncio, Herley and Coskun ask "Do strong web passwords accomplish anything?" \cite{florencio2007strong}. They suggest that strength rules for web passwords accomplish very little when a lockout rule can restrict access. Beautement et al. \cite{beautement2009compliance} introduce the idea of a compliance budget which formalizes the understanding that users and organizations do not have unlimited capacity to follow new instructions and advice. This is an important concept for us to keep in mind when we look at the costs of implementing advice. 

In this paper, Section \ref{sec:col} will explain how we collected the password advice and Section \ref{sec:cat} illustrates the steps we took for categorizing it. Section \ref{sec:idcosts} describes our methods for identifying costs and how we assigned preliminary costs to a subset of the advice collected. Lastly, Section \ref{sec:disc} discusses the characteristics associated with password advice which were highlighted during this process. 

\section{Collection of Advice} \label{sec:col}
To begin studying password advice, we first needed to collect a selection of the advice that is distributed to users. We primarily used Internet searches to collect password advice but also looked at advice given by standards agencies and multinational companies. We attempted to recreate the actions an individual or organization might take when seeking to inform themselves about proper password practices. As such, while the advice given in academic papers might be more considered, if it was not easily accessible, we did not include it in our study. In total, we collected 269 pieces of password advice from 21 different sources. Table \ref{tab:Sources} shows the types of sources from which the advice was gathered. 

\begin{table}[htb]
\caption{\label{tab:Sources}Break down of advice sources.}
\small
\centering
\begin{tabular}{|l|l|}
\hline
Source&Number\\\hline
Multinational companies&6\\
Universities&6\\
Security specialists&5\\
General articles&4\\\hline
\end{tabular}
\end{table}
\FloatBarrier
\section{Categorizing Advice} \label{sec:cat}
To extract meaning from the pieces of advice that we collected we subdivided the advice into categories. Within each category we created statements that generalized the recommendations pertaining to each category. 

\subsection{Categorization}
Our first step after collecting the advice was to group it into categories. For this we considered each piece of advice individually. The first pieces of advice we examined suggested our starting categories. From there, each piece of advice was either included in one of our existing categories, expanded the scope of an existing category or created a new category to suit it. For example, when approaching a piece of advice which said "Use a unique password for each of your important accounts" we created the category \textit{Reuse across Accounts}. However, when a second piece of advice stated "Don’t recycle passwords" we altered the name of the category to be the more general \textit{Password reuse}. As an example, in Figure \ref{fig:Reuse2} we show the seventeen pieces of advice which became grouped under the category \textit{Password reuse}. 

In total, we identified 29 categories shown in Table \ref{tab:Categories}. The categories are listed in two columns; one showing categories containing advice aimed at users and the second showing advice aimed primarily towards organizations. Also included are the number of pieces of advice under each category. In this paper we only have space to include analysis of four of these categories. They are shown in italics in Table \ref{tab:Categories}. 

We collected 155 pieces of advice aimed towards users and 114 pieces aimed towards organizations. Despite its greater quantity user advice has been subdivided into fewer categories. We speculate this could be related to the wider variety of roles an organization plays in the security of passwords. But it could also reflect our greater familiarity with user advice. While everyone is a user, not everyone has held all roles in an organization and therefore it was easier to interpret and categorize user advice. 

\begin{table}
\caption{\label{tab:Categories}Categories and the quantity of advice they contain.}
\small
{\renewcommand{\arraystretch}{1}%
\begin{tabular}{|l|c|l|c|}\hline
\rule{0pt}{1em}\textbf{Users}&\#&\textbf{Organisations}&\#\\\hline
\rule{0pt}{1em}\textit{Phrases}&37&\textit{Expiry}&27\\
\textit{Composition}&28&Length&17\\
Personal Information&21& Storage&13\\
\textit{Reuse}&17&Keeping system safe&8\\
Personal pwd storage&17&Throttling guesses&8\\
Backup pwd options&8&Individual accounts&7\\
Sharing&14&Generated pwds&6\\
Keeping account safe&8&Transmitting pwds&2\\
Password managers&4&Admin accounts&4\\
Username requirements&2&Default passwords&4\\
Two step verification&1&Shoulder surfing&3\\
Two factor authentication&2&Access to pwd file&3\\
&&Policies&2\\
&&Input&3\\
&&Network strings&2\\
&&Cracking&1\\
&&Back up work&1\\\hline
\rule{0pt}{1em}\textbf{Total}&\textbf{159}&\textbf{Total}&\textbf{111}\\\hline
\end{tabular}}
\end{table}
 
\subsection{Classification into statements}
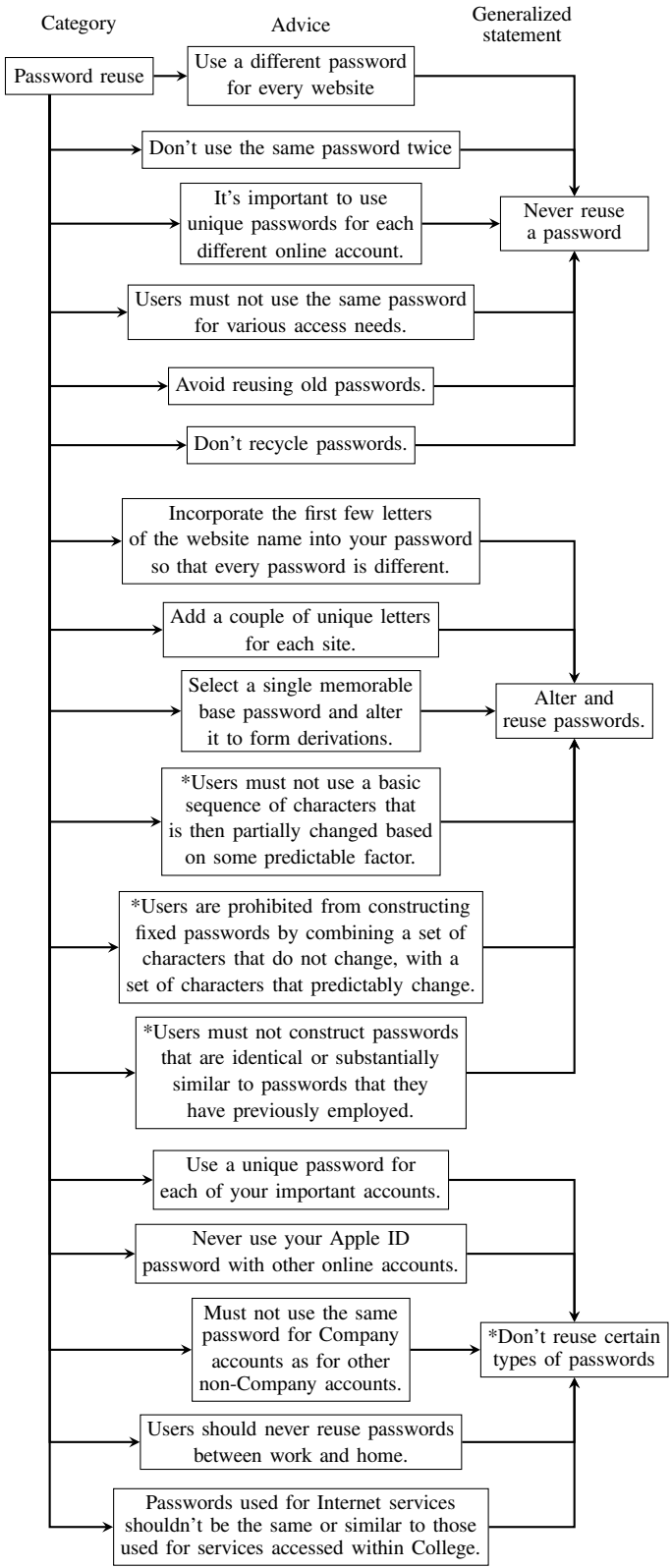
\begin{figure}
\footnotesize
\centering
\tikzstyle{block} = [rectangle, minimum width=2cm, minimum height=0.5cm,text centered, draw=black]
\tikzstyle{heading} = [rectangle, minimum width=2cm, minimum height=0.5cm,text centered]
\tikzstyle{arrow} = [thick,->,>=stealth]

\begin{tikzpicture}[auto, node distance=5cm,>=latex']

\node (co3) [block] {Password reuse};
\node (co2) [coordinate, xshift=-0.4cm, yshift=-0.25cm] {};

\node (h1) [heading, above of=co3, yshift=-4.3cm, xshift=0] {Category};
\node (h2) [heading, right of=h1, xshift=-2cm] {Advice};
\node (h3) [heading, right of=h2, xshift=-2cm] {\shortstack{Generalized\\statement}};

\node (in1) [block, right of=co3, xshift=-2cm] {\shortstack{Use a different password\\ for every website}};
\node (in2) [block, below of=in1, yshift=4cm] {Don't use the same password twice};
\node (in3) [block, below of=in2, yshift=4cm] {\shortstack{It's important to use\\unique passwords for each\\different online account.}};
\node (in4) [block, below of=in3, yshift=3.8cm] {\shortstack{Users must not use the same password\\ for various access needs.}};
\node (in5) [block, below of=in4, yshift=4cm] {\shortstack{Avoid reusing old passwords.}};
\node (in6) [block, below of=in5, yshift=4.2cm] {\shortstack{Don’t recycle passwords.}};


\node (in7) [block, below of=in6, yshift=3.7cm] {\shortstack{Incorporate the first few letters \\of the website name into your password \\so that every password is different.}};
\node (in8) [block, below of=in7, yshift=3.8cm] {\shortstack{Add a couple of unique letters\\ for each site.}};
\node (in9) [block, below of=in8, yshift=3.9cm] {\shortstack{Select a single memorable\\ base password and alter\\ it to form derivations.}};
\node (in10) [block, below of=in9, yshift=3.5cm] {\shortstack{*Users must not use a basic\\ sequence of characters that\\ is then partially changed based\\on some predictable factor.}};
\node (in11) [block, below of=in10, yshift=3.3cm] {\shortstack{*Users are prohibited from constructing\\ fixed passwords by combining a set of\\ characters that do not change, with a\\set of characters that predictably change.}};
\node (in12) [block, below of=in11, yshift=3.3cm] {\shortstack{*Users must not construct passwords\\ that are identical or substantially\\ similar to passwords that they\\ have previously employed.}};

\node (in13) [block, below of=in12, yshift=3.55cm] {\shortstack{Use a unique password for \\each of your important accounts.}};
\node (in14) [block, below of=in13, yshift=4cm] {\shortstack{Never use your Apple ID\\ password with other online accounts.}};
\node (in15) [block, below of=in14, yshift=3.7cm] {\shortstack{Must not use the same\\password for Company\\accounts as for other\\non-Company accounts.}};
\node (in16) [block, below of=in15, yshift=3.75cm] {\shortstack{Users should never reuse passwords \\between work and home.}};
\node (in17) [block, below of=in16, yshift=3.85cm] {\shortstack{Passwords used for Internet services\\shouldn't be the same or similar to those\\used for services accessed within College.}};

\node (out1) [block, right of=in3, xshift=-1.3cm] {\shortstack{Never reuse\\a password}};
\node (out2) [block, right of=in9, xshift=-1.3cm] {\shortstack{Alter and\\reuse passwords.}};
\node (out3) [block, right of=in15, xshift=-1.3cm] {\shortstack{*Don't reuse certain\\types of passwords}};

\draw [arrow] (co3) -- (in1);
\draw [arrow] (co2) |- (in2);
\draw [arrow] (co2) |- (in3);
\draw [arrow] (co2) |- (in4);
\draw [arrow] (co2) |- (in5);
\draw [arrow] (co2) |- (in6);
\draw [arrow] (co2) |- (in7);
\draw [arrow] (co2) |- (in8);
\draw [arrow] (co2) |- (in9);
\draw [arrow] (co2) |- (in10);
\draw [arrow] (co2) |- (in11);
\draw [arrow] (co2) |- (in12);
\draw [arrow] (co2) |- (in13);
\draw [arrow] (co2) |- (in14);
\draw [arrow] (co2) |- (in15);
\draw [arrow] (co2) |- (in16);
\draw [arrow] (co2) |- (in17);

\draw [arrow] (in1) -| (out1);
\draw [arrow] (in2) -| (out1);
\draw [arrow] (in3) -- (out1);
\draw [arrow] (in4) -| (out1);
\draw [arrow] (in5) -| (out1);
\draw [arrow] (in6) -| (out1);
\draw [arrow] (in7) -| (out2);
\draw [arrow] (in8) -| (out2);
\draw [arrow] (in9) -- (out2);
\draw [arrow] (in10) -| (out2);
\draw [arrow] (in11) -| (out2);
\draw [arrow] (in12) -| (out2);
\draw [arrow] (in13) -| (out3);
\draw [arrow] (in14) -| (out3);
\draw [arrow] (in15) -- (out3);
\draw [arrow] (in16) -| (out3);
\draw [arrow] (in17) -| (out3);

\end{tikzpicture}
\caption{Method for categorizing advice. Example: Reuse.} \label{fig:Reuse2}
\end{figure}

Once we divided the advice into categories we noticed the pieces of advice within each category did not necessarily promulgate similar opinions. It was therefore necessary to subdivide the advice into statements which offer a similar message. In Figure \ref{fig:Reuse2} we can see how the seventeen pieces of advice under \textit{Password Reuse} were grouped into three distinct statements:
\begin{itemize}
\item Never reuse a password.
\item Alter and reuse passwords.
\item Don't reuse certain passwords. 
\end{itemize}
In this figure we make note of pieces of advice that contradict the main statement with a star (*). It is important to note that while it appears that there is no contradictory advice within the category statement "Never reuse a password" the third statement "Reuse certain passwords" is itself a contradiction. We make note of this by placing a star (*) in the text box of the third category. It is also represented by a star in Table \ref{tab:costs}. Already we are beginning to see inconsistencies with the advice that is circulated.

Thus, within each category we created generalized statements of the advice that was given. It is worth noting that the labels for advice are given from the perspective of the majority. For example, if two pieces of advice state that passwords should not include published phrases and one piece of advice states that it would be a good idea to use published phrases then the advice will be labeled as "Don't include published phrases". 

The statements relating to the four chosen categories are shown on the left hand side of Table \ref{tab:costs} on page \pageref{tab:costs}. For each statement the table shows how many pieces of the advice agree with the sentiment of the statement and how many disagree. This gives a clear indication of the inconsistencies in circulated password advice. 

\section{Identification of Costs} \label{sec:idcosts}
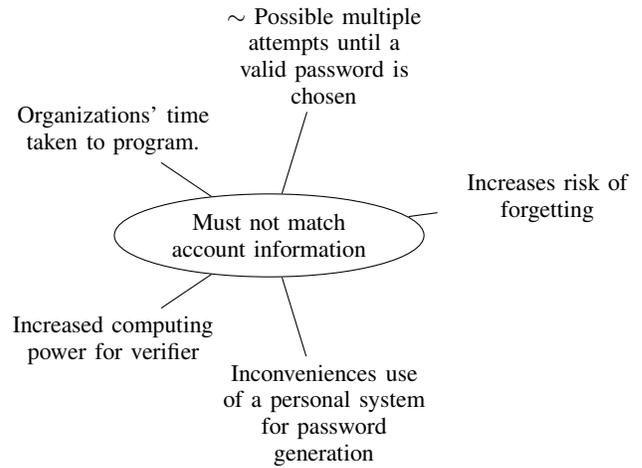
\begin{figure}
\small 

\tikzstyle{block} = [ellipse, minimum width=1cm, minimum height=0.8cm,text centered, draw=black]
\begin{tikzpicture}[grow cyclic, text width=2.7cm, align=flush center,
    level 1/.style={level distance=2.5cm,sibling angle=73}]

\node[block]{Must not match account information}
     	child { node {Increased computing power for verifier}}
   		child { node {Inconveniences use of a personal system for password generation}}
    	child [xshift=1.2cm, yshift=0.5cm]{ node {Increases risk of forgetting}}
        child { node {$\sim$ Possible multiple attempts until a valid password is chosen}}
        child { node {Organizations' time taken to program.}}
    ;
   
\end{tikzpicture}

\caption{Identifying costs for the statement "Must not match account information".} \label{fig:MatchAcc}
\end{figure}

As described in Section \ref{sec:cat}, we categorized the advice collected into 29 categories and 78 statements. For each statement, we identified the costs we believe are associated with it. Figure \ref{fig:MatchAcc} shows an example for the statement "Passwords must not match account information". 

We did not restrict ourselves in the types of costs we identified. In this way, we were not limiting ourselves by trying to stay within a predetermined structure. Despite this, we nearly immediately saw similarities in the costs we were identifying. 
After analyzing each of the 78 statements we had identified 10 categories of costs that we believe provide a rudimentary understanding of the general costs associated with obeying password advice. These are shown in Table \ref{tab:costs categories}.

\begin{table}
\small
\caption{\label{tab:costs categories}Cost categories.}
\centering
\begin{tabular}{|c|p{7cm}|}\hline
\rule{0pt}{1em}&Costs\\\hline
\rule{0pt}{1em}1.&Increased risk of forgetting.\\
2.&Need to pick a new password.\\
3.&Possible multiple attempts needed to enter a valid password.\\
4.&Inconveniences use of a personal system for password generation.\\
5.&User time taken.\\
6.&Reduced "entropy".\\
7.&Organizations' time taken to enforce/program.\\
8.&Impossible/hard to enforce.\\
9.&Creates an additional security hole.\\
10.&Increased computing power needed.\\\hline
\end{tabular}
\end{table}

\subsection{Discussion of cost categories}
When determining categories of costs we noticed that some categories are outcomes of others, similarly, many categories can be seen to have "sub-costs". For example, \textit{increased risk of resets} is a sub-cost of \textit{increased risk of forgetting}. Similarly, \textit{abandon site} can be seen as a result of \textit{Possible multiple attempts needed to enter a valid password} and \textit{irritating} \cite{passwordfatigue} can be seen as an outcome of many of the other categories. In order to minimize this, we reduced the number of categories in the knowledge that we believe most categories will inevitably contain sub-costs. Note that we do keep \textit{user time} as a distinct category as at times we recognize that there is no cost other than the users' time. 

One of the cost categories we created was \textit{Reduced "entropy"}. We are defining this to be a reduction in the number of guesses an attacker needs to make or a reduction in the keyspace or entropy (measure of uncertainty) \cite{pliam1998disparity}\cite{bonneau2012science}. We are not considering guesswork to be a substitute for entropy \cite{malone2005guesswork} but using the word "entropy" as a general word for guessability, keyspace and entropy. This is because relying solely on entropy oversimplifies how passwords withstand guessing attacks \cite{zhang2016revisiting}. In many cases, the trade-off between keyspace/entropy/guessability is not clear and requires more information for a definitive answer. 

We have presented the costs for the four selected categories in Table \ref{tab:costs}. Each scheme is rated as either containing the cost (\newmoon) or not (no-circle); if a scheme contains some part of the cost or some variation of the cost, we use the "Quasi-" prefix (\fullmoon) to indicate this. This is inspired by the system created by Bonneau et al. in their framework for analyzing password alternatives \cite{bonneau2012quest}. 


\begin{table*}[ht!]
\centering
\small
\caption{\label{tab:costs}Costs of implementing password advice.}
\begin{supertabular}{lrr|c|c|c|c|c|c|c|c|c|c|} \cline{4-13}

\multicolumn{1}{c}{} &\multicolumn{2}{c}{} & \multicolumn{10}{|c|}{\rule{0pt}{1em}Costs}\\\cline{4-13}
&\rot{\#\ Against} & \rot{\#\ Supporting} & \rot{\shortstack{Increased risk of\\forgetting}} & \rot{\shortstack{Need to pick a new\\password}} & \rot{\shortstack{Possible multiple\\attempts needed}} & \rot{\shortstack{Inconveniences use of\hphantom{v}\\personal system for\\password generation}} & \rot{User time} & \rot{Reduced "entropy"} 
        & \rot{\shortstack{Organizations' time\\to enforce/program}} & \rot{\shortstack{Impossible/hard\\to enforce}}  & \rot{\shortstack{Creates an additional\\security hole}} & \rot{\shortstack{Increased computing\\power needed}} \\\hline

\multicolumn{1}{l}{Phrases} &  & \multicolumn{1}{c|}{} & \multicolumn{9}{c}{}& \multicolumn{1}{c|}{} \\\hline
Don't use patterns.  &0 & 6 & \newmoon &  & \newmoon & \newmoon &  &  & \newmoon& \fullmoon & &   \\
Take initials of a phrase. &0 & 4 & \fullmoon &  &  &  &  &  &  &\newmoon &  &   \\
Don't use published phrases. & 1 & 2 & \newmoon &  &  & \fullmoon &  &  & \newmoon & \fullmoon &  &   \\
Substitute symbols for letters. &1 & 2 & \newmoon &  &  &  &  &  &  & \newmoon &  &  \\
Don't use words.  &0 & 16 & \newmoon &  & \newmoon & \newmoon &  & & \newmoon & &  &  \\\hline

\multicolumn{1}{l}{Composition} & &  \multicolumn{1}{c|}{} & \multicolumn{9}{c}{} & \multicolumn{1}{c|}{}\\\hline
Must include special characters  & 5 & 7 &\newmoon & & \fullmoon &  &  & \fullmoon& \newmoon &  &  &   \\
Don't repeat characters. & 0&  3 & \newmoon &  &\fullmoon & \newmoon &  & \newmoon & \newmoon &  & &  \\
Enforce restrictions on characters.  & 1 & 12 & \newmoon & & \newmoon & \newmoon &  & \fullmoon &  &  &  &  \\\hline

\multicolumn{1}{l}{Expiry} &  & \multicolumn{1}{c|}{} & \multicolumn{9}{c}{}& \multicolumn{1}{c|}{} \\\hline
Store history to eliminate reuse.  &0 &5  & \fullmoon & \newmoon& \newmoon &  &  &  & \newmoon&  & \newmoon &\newmoon  \\
Have a minimum Password Age.&0 &1  &  &  &  &  &  & &\fullmoon &  & \fullmoon &   \\
Change your password regularly.&4 &7  & \newmoon& \newmoon &  &  & \newmoon & \fullmoon & \newmoon &  &  &   \\
Change if suspect compromise.&0 &10  &  & \newmoon &  & &  &  & \fullmoon &  &  & \fullmoon  \\\hline

\multicolumn{1}{l}{Reuse} & & \multicolumn{1}{c|}{} & \multicolumn{9}{c}{} &\multicolumn{1}{c|}{} \\\hline
Never reuse a password. &*5& 6 & \newmoon & \newmoon &  &  &  &  &  & \newmoon &  & \\
Alter and reuse passwords& 3 &  3 & \fullmoon &  &  &  &  &  &  & \newmoon & \fullmoon &  \\
Don't reuse certain sites' passwords.  & 0 & 5 &  &  &  & &  &  &  & \newmoon & \fullmoon &   \\\hline

\end{supertabular}

\raggedright \begin{center} \newmoon  = requires the cost; \fullmoon  = partially requires the cost; no circle = does not require the cost. \vspace{-2mm}\end{center} 
\end{table*}

\section{Discussion} \label{sec:disc}
We will discuss some observations that we made for the chosen subset of advice categories. Depending on the category we will either discuss each statement in turn or consider the statements where the advice is unanimous and then the statements for which the advice is contradictory.

\subsection{Phrases}
Advice regarding password phrases was the most commonly given advice we encountered. This implies that advice is mostly concerned with making passwords "strong". Yet in some cases, the strength of a password is irrelevant to defend users, as with password capture (e.g. phishing, keylogging) \cite{zhang2016revisiting}. In fact, Bellovin 2008 \cite{bellovin2008security} claims that the most common way passwords are compromised is via keylogger attacks.


\subsubsection{Unanimous}
Within the category \textit{Phrases} there were no contradictions for the statements: \textit{Don't use patterns, Take initials of a phrase} and \textit{Don't use words.} The last is particularly interesting since from leaked password database we know users primarily chose word based passwords \cite{rockyou}. Shay et al. find that the "use of dictionary words and names are still the most common strategies for creating passwords" \cite{shay2010encountering}. This depicts how ineffective some password advice can be and is possibly a reflection on the costs appearing to not outweigh the benefits from a users' point of view. 

\subsubsection{Contradicting}
The statements: \textit{Don't use published phrases} and \textit{Substitute symbols for letters} had contradictions. For \textit{don't use published phrases} the advice given was:
\begin{enumerate}
\item"Don’t use song lyrics, quotes or anything else that has been published."
\item"Do not choose names from popular culture."
\item"Choose a line of a song that other people would not associate with you."
\end{enumerate}
The last piece of advice directly contradicts the first. This type of inconsistency in the advice given makes it no surprise that users seem disinclined to follow security advice \cite{inglesant2010true}\cite{adams1999users}.

The advice statement \textit{Substitute symbols for letters} is proposed by two sources but is advised against by a third. We know from Warner \cite{subs} that passwords with simple character substitutions are weak. Yet, 2 of 3 pieces of advice recommend it. This could stem from the attitude that it is "better than nothing" and, as we can see from Table \ref{tab:costs}, the cost to the user is relatively low. 

\subsection{Composition}
Composition restrictions are regularly enforced by websites but the advice relating to this is not consistent from site to site. It is interesting to note that Herley \cite{herley2009so} hypothesizes that different websites may deliberately have policies which are restrictive to different degrees. As this can help ensure that users do not share passwords between sites. Below we will discuss each of the three statements associated with composition. 

\subsubsection{Must include special characters}
Seven sites instructed users to \textit{include special characters} in their passwords, but five sites placed restrictions on the special characters that could be used. The main restriction on special characters was "do not use spaces". However, one piece of advice stated the more direct "do not use special characters". By not allowing users to include all special characters an attackers' search space is decreased. 

\subsubsection{Don't repeat characters}
Not allowing the repetition of characters deters users from choosing passwords such as "aaaaaaa" or "wwddcc". Depending on the strictness of the restriction it could eliminate words such as "bookkeeper" or "goddessship". It could also cause some inconvenience for random password generators where the word "Sdt2htTtd65c8h" could be rejected. We list it as incurring the cost \textit{reduced "entropy"} since it is banning characters sequences.

\subsubsection{Enforce restrictions on characters}
We collected twelve pieces of advice encouraging composition restrictions on passwords and only one piece of advice against it. The source rejecting composition rules was the NIST 2016 draft password guidelines. Though the guidelines are still in the review stage they are receiving promising responses from the research community \cite{nistcox}. This raises the question: will organizations begin to disseminate these new security practices? Or continue to enforce their stringent password restrictions?

We claim that forcing users to include special characters\textit{ quasi-reduces "entropy"}. If a user creates an eight digit passwords with no restrictions each of the eight characters could be any of the 96 possible ASCII characters. However, by restricting the password so that it must include one special character we limit the options for one of the character to the 34 special characters. This becomes more significant when a site enforces multiple restrictions on composition. In addition, the probability of a user including a "1" as their number and an "!" as there symbol is high \cite{ur2015added}. So again an attacker can refine the guesses they make. This idea of composition restrictions reducing search space is something we will consider further in future work.

\subsection{Expiry}
\subsubsection{Unanimous}
We found five pieces of advice telling organizations to \textit{Store password history to eliminate reuse}, one encouraging organizations to \textit{Enforce a minimum password age} and ten in favor of \textit{Changing passwords if compromise is suspected.}
If organizations do \textit{store their users' password history} this \textit{creates an additional security hole} as the company needs to allocate resources to protecting this file. Users can no longer reuse prior passwords but alterations are still possible \cite{shay2010encountering}. In fact, Zhang, Monrose and Reiter \cite{zhang2010security} identify that we can easily predict new passwords from old when password aging policies force updates.

The reason given for introducing a \textit{minimum password age} is to prevent users from bypassing the password expiry system by entering a new password and then changing it right back to the old one \cite{technetmag}. However, if an attacker gains access to a users' account and changes their password the user will be unable to change it again until the required number of days have elapsed, or with an administrators' help. 

Ten pieces of advice recommended \textit{changing passwords if a compromise is suspected}. This can be inconvenient for users not affected by the compromise, and also those that are. If there is a breach at the server the users were not at fault, yet still they must choose a new password. 

\subsubsection{Contradicting} 
From anecdotal evidence we know the advice \textit{change your password regularly} is widely hated by users \cite{hatechangingpass}. Referring to the costs in Table \ref{tab:costs}, we note that the costs associated with other advice are one-time occurrences. By contrast, when password expiry is enforced users face many of the costs periodically. Seven pieces of the advice we collected encouraged the use of password expiry while only four pieces of advice discouraged it. This is despite research suggesting that the security benefits are minimal \cite{chiasson2015quantifying}\cite{zhang2010security}. This implies the inconvenience to users is worth less to organizations than the minimal security benefits. Or do organizations want to be seen to be enforcing strong security practices, and forcing expiry is just one way of doing this?

\subsection{Reuse}
We collected six pieces of advice telling users to \textit{never reuse passwords} and three pieces telling users to \textit{not reuse passwords for certain sites}. In addition, we found three pieces of advice encouraging users to \textit{alter and reuse their passwords} and three pieces telling users to not alter and reuse their passwords. There seems to be little agreement among the distributed advice in terms of password reuse. 

\subsubsection{Never reuse a password vs. reuse for certain accounts}
Das et al. estimate that 43-51\% of users re-use passwords across sites \cite{das2014tangled}. They also provide algorithms that improve an attacker’s ability to exploit this fact. Flor{\^e}ncio, Herley and Van Oorschot \cite{florencio2014password} declare that, far from being unallowable, password reuse is a necessary and sensible tool for managing a portfolio of passwords. They recommend grouping passwords according to their importance and reusing passwords only within those groups. Interestingly, the advice we collected \textit{Don't reuse certain passwords} gave a slightly different take on this advice. The advice mostly asked users to not use the password used for their site anywhere else e.g. "Never use your Apple ID password for other online accounts". Most organizations gave advice prioritizing their own accounts. Only one piece of advice suggested using a unique password for any important accounts \cite{google}.

\subsubsection{Alter and reuse passwords}
An alternative to grouping accounts for reuse is to alter and then reuse a password. This advice was given by three sources and rejected by three sources. These alterations are sometimes very predictable. Using a cross-site password guessing
algorithm Das et al. \cite{das2014tangled} were able to guess approximately 10\% of non-identical password pairs in less than 10 attempts and approximately 30\% in less than 100 attempts. We could find no research identifying this method of altering and reusing passwords as effective. We consider altering and reusing passwords to \textit{quasi-increase the risk of forgetting}, \textit{impossible to enforce} and \textit{quasi-creates an additional security hole}.

\FloatBarrier

\section{Conclusions}
In this paper, we highlighted characteristics of the password advice currently available online. We show that there are serious discrepancies in the advice given between sources. We also note that some of the advice viewed by researchers and specialists as "best practice" is often not represented by the majority of advice. This contradictory information may reflect one of the reasons for users' unwillingness to follow advice. 

We then looked at costs that could be associated with the enforcement of different pieces of password advice. Our aim here was to introduce the idea of a framework for deducing costs associated with implementing password advice. The costs model also provides some rudimentary insight into the biases of password advice.

Our collection and categorization of advice and identification of costs brought discrepancies in password advice into focus, in addition it highlighted the following interesting characteristics:

Research has shown that substituting symbols for letters is a weak security practice. But two of three pieces of advice recommend it.

Most of the advice we collected was concerned with making passwords strong. Yet, password strength cannot protect against password capturing malware, social engineering, or physical observation.

Of the 13 pieces of advice relating to password composition only the NIST 2016 draft guidelines spoke against restrictions.

Sixteen pieces of advice recommended that words not be included in passwords but we know from leaked password databases that users primarily choose word based passwords.

Seven pieces of the advice we collected encouraged the use of expiration policies while four discouraged it. The costs associated with expiry imply that the inconvenience to users is worth less to an organization that the minimal security benefits. 

Similarly, most organizations gave advice encouraging the prioritization of passwords associated with their own accounts rather than encouraging realistic and user-focused security practices. 

In terms of future work, our next step is to develop methods to quantify each of the costs we have identified in this paper.

\bibliographystyle{ieeetr}
\bibliography{library.bib}

\end{document}